\def\imat {i}

\documentclass[prb,howpacs,twocolumn,floats,epsf,ifthen,aps]{revtex4}

\usepackage{graphicx}
\usepackage{amsmath}

\begin{document}

\title{Manipulation of double-dot spin qubit by continuous noisy measurement}

\author{Rusko Ruskov, %\footnote{On leave from INRNE, Sofia BG-1784, Bulgaria}
%\footnote{On leave of absence from Institute
%for Nuclear Research and Nuclear Energy, Sofia BG-1784, Bulgaria}
Viatcheslav V. Dobrovitski, and Bruce N. Harmon}

\address{Ames Laboratory, Iowa State University, Ames, Iowa 50014, U.S.A.}

\date{\today}

\begin{abstract}
We consider evolution of a double quantum dot (DQD) two-electron
spin qubit which is continuously weakly measured with a linear charge
detector (quantum point contact). Since the interaction between
the spins of two electrons depends on their charge state, the
charge measurement affects the state of two spins, and induces
non-trivial spin dynamics. We consider the regimes of strong and weak
coupling to the detector, and investigate the measurement-induced
spin dynamics both analytically and numerically.
We observe emergence of the negative-result evolution and the
system stabilization due to an analog of quantum Zeno effect.
Moreover, unitary evolution between the triplet and
a singlet state is induced by the negative-result measurement.
We demonstrate that these effects exist for both strong and
weak coupling between the detector and the DQD system.
\end{abstract}

%\pacs{03.65.Ta, 85.25.Dq, 03.65.Yz }
\maketitle

\section{Introduction}

Quantum dots (QDs) recently have attracted much attention as very
suitable candidates for studying evolution of a single quantum
system. Moreover, QDs are very promising candidates for quantum
information processing. An electron spin in a quantum dot is a good
representation for a qubit, being a natural two-state quantum
system. Manipulations of the single QDs and double-QD (DQD) systems
can be achieved using %the
external magnetic fields and gate voltages
\cite{HansonReview}, which implement injection of an electron (or
pair of electrons for DQD) in a given spin state, or some unitary
rotation in a two-spin space. However, it is not easy to achieve the
full set of transformations in the Hilbert space of two coupled
spins \cite{QDman1,QDman2,QDman3,QDman4}: some transformations
require  a
strong gradient of magnetic fields on a nanometer scale, and
advanced techniques are needed for performing spin rotations rapidly
and reliably.

Thus, it is interesting to explore other possible ways of manipulating
electron spins in DQD structures. In particular, it is natural
to study whether measurement of the charge state of the DQD may
help drive %ing
the desired evolution of the two-spin system.
The weak continuous measurement \cite{Dalibard,Carmichael,WisemanMilburn,Kor-99-01},
which monitors the system in question, and therefore affects its evolution,
may provide an additional and useful tool for controlling the
electron spins in DQD systems.
At some level of description of a probed quantum system plus detector,
it is postulated that the fundamental measurement process consists of
direct particle detections (e.g., photons, electrons, etc.)
or absence of any detection.
The absence of a detection for specific time interval constitutes
a negative result. In both cases (``detection'' or ``no detection'')
the quantum
state of the system evolves according to the provided information
and the evolution can be
%interpreted as a result of
described using quantum Bayesian inference.\cite{Kor-99-01}
%,Gardiner}
E.g.,
for a two-level atom which did not decay by time $t$
the evolution of the atom's density matrix takes the form:
\begin{eqnarray}
&&\rho_{ee}( t) = \frac{\rho_{ee}(0)\, e^{-\Gamma_{sp}t}}{\cal N} ,\
\rho_{gg}( t) = \frac{\rho_{gg}(0)}{\cal N}  \ \
    \label{atom-Bayes1}\\
&& |\rho_{eg}( t)| = |\rho_{eg}(0)|\,
\sqrt{\frac{\rho_{ee}(t)\,\rho_{gg}(t)}{\rho_{ee}(0)\,\rho_{gg}(0)}}  , \qquad
\label{atom-Bayes2}
\end{eqnarray}
where $e^{-\Gamma_{sp}t}$
is the conditional probability not to decay from the excited state
$|e\rangle$ (ground state $|g\rangle$ is supposed stable),
$\Gamma_{sp}$ is the spontaneous decay rate, and ${\cal N}\equiv
\rho_{gg}(0) + \rho_{ee}(0)\, e^{-\Gamma_{sp}t}$.
The classical Bayesian update of the diagonal
elements, Eq.(\ref{atom-Bayes1}), reflects an information-related
evolution.
Also, Eq.(\ref{atom-Bayes2}) shows that the coherence ratio
$|\rho_{eg}|/\sqrt{\rho_{ee}\,\rho_{gg}}$ is conserved;
in particular, a pure state remains pure.

The above evolution has been proposed and discussed mainly
in the context of quantum optics.
\cite{Epstein-Dicke}${}^{,}$\cite{Dalibard,Carmichael,WisemanMilburn}
For solid state systems,
it has been realized recently, using an artificial two-level system:
a superconducting ``phase'' qubit, measured via tunneling of
its superconducting phase.\cite{Katz-Sci}
The detection of tunneling
was considered as an instantaneous process\cite{Katz-Sci,Pryadko}
in the highly non-linear detector (dc-SQUID)
that implies detector relaxation rates
much faster than
the qubit ``decay rate''.
\begin{figure}
\vspace*{0.1cm}
\centering
\includegraphics[width=2.8in]{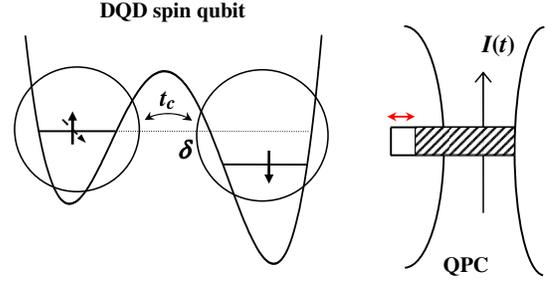}
\vspace{-0.06cm}
\caption{Schematic of a DQD spin system measured by a QPC;
the QPC current $I(t)$ is affected by the charge state of the DQD.
In quantum Zeno regime the tunneling between dots is suppressed
while the spin qubit undergoes continuous negative-result evolution.}
\label{schematic}
\end{figure}
In this paper we consider
another realization of a solid state
qubit and a linear detector measuring continuously.
Nevertheless, it can behave similarly to the above examples in
certain measurement regimes.
We will consider a semiconductor double quantum dot (DQD) spin
system (Fig.\ref{schematic}) where two electrons can be in either of
the dots and the relevant states are spin-charge
states.\cite{LossDiVi,Kouwenhoven-Nat1,Marcus-Sci,HansonReview}
At relatively small time scales (from ns to $\mu$s)
the electron tunneling between the dots preserves the
system's total spin
that allows spin-selective evolution and
leads to the so called spin-to-charge conversion.\cite{HansonReview}
This allows the measurement of the DQD spin system by a charge
sensitive detector such as a quantum point contact
(QPC)\cite{Kouw-prl94,Marcus-Sci} or single electron transistor
(SET).\cite{Rimberg-nat}

For our system (Fig.\ref{schematic})
the spin qubit\cite{Marcus-Sci} is formed by
the singlet spin state $|S(1,1)\rangle$ and the triplet state
$|T_0(1,1)\rangle$
that has a zero spin projection on an external
magnetic field
(here, e.g., $(1,1)$ denotes a charge state when
one electron is in the left dot and the other is in the right dot).
Spin conservation allows transition between the singlet $|S(1,1)\rangle$
and the localized singlet, $|S(0,2)\rangle$, while a transition
from the triplet state is spin-blocked.
Thus, the state $|S(0,2)\rangle$ can be used
for preparation and subsequent measurement of the spin qubit.
For a weak measurement the system-detector coupling $\Gamma$ is finite and
defines the time scale of measurement evolution
(in particular, $1/\Gamma$ is of the order of the time needed
to distinguish the two different charge states by the QPC).
The spin system evolution will be an interplay
of the non-unitary dynamics due to measurement and the
dynamics associated with the system's
internal Hamiltonian.

The purpose of this work is to develop the theory of quantum evolution
of a DQD two-electron spin system under the
continuous measurement by a linear charge detector  such as a QPC
that is based on the Bayesian approach.\cite{Kor-99-01,Kor-osc,Ruskov-neg}
Initially motivated by
experiment\cite{Marcus-Sci}
we pose the question:
``How the measurement by a QPC will affect the spin qubit state
and is it possible to use continuous measurement for manipulation of the
quantum state ?''
The system state is continuously updated according to a given measurement record
due to quantum back action.\cite{Kor-99-01}
On typical system's times
when many electrons pass through the QPC, the quantum
evolution
is conditioned on the fluctuating  detector current $I(t)$.

We identify regimes when the system-detector coupling
becomes large relative to a typical system frequency,
that leads to
suppression of coherent tunneling transitions.
The associated stabilization of the states in continuous measurement
(see, e.g., Ref.\cite{Kor-osc,RusKor-ent,Ruskov-Zeno})
is an analog of the well known quantum Zeno effect.\cite{Khalfin,MisraSudarshan,Peres}
Since the system-detector coupling is finite, the Zeno stabilization
does not last forever.
The stabilized states switch to each other
with a rate $\Gamma_{sw} \ll \Gamma$.
The switching
happens on a time scale $1/\Gamma$.
In fact, on this time scale the
switching transition becomes irreversible.
For an ensemble averaged description of the state evolution
we will show that the irreversibility corresponds to an exponential decay
(compare with Ref.\cite{Pryadko})
with the decay rate $\Gamma_{sw}$.

Correspondingly, in the case of  a  spin qubit
we show via numerical simulation of the measurement process
that the switching rate $\Gamma_{sw}$ plays the role of the
spontaneous decay rate $\Gamma_{sp}$ for the two-level atom described above.
Despite the quantum back action noise reflecting the noisy output,
the spin qubit state will evolve according to
Eqs.(\ref{atom-Bayes1}), (\ref{atom-Bayes2}) in a spin-charge basis
that corresponds to the Zeno stabilized states.
In particular, in our analysis the negative-result evolution of
the qubit subsystem is not postulated  but  emerges from
continuous noisy measurement evolution and Hamiltonian evolution
at a microscopic level.

The paper is organized as follows.
In Sec.\,II we review/derive the Bayesian equations
for weak continuous measurement of a DQD two-electron spin system.
In Sec.\, III we investigate
the measurement collapse scenarios
for different coupling regimes
and give qualitative and quantitative explanations of the obtained results.
In Sec.\, IV we estimate the detector-system coupling and other important
parameters that could be relevant for an experiment to confirm our results.
We also comment on the role of various sources of decoherence.

\section{DQD two electron system and measurement device}

\subsection{Hamiltonian and time scales}
\label{Hamiltonian_time_scales}

The system of two electrons confined in the DQD, and the quantum point contact
as a charge sensing detector are shown in Fig.\ref{schematic}.
External gate voltages form the potential profile of the DQD
and are well controlled in the experiment.\cite{HansonReview}
In particular, changing the energy difference between the dots, $\delta$,
allows continuous tuning of
the charge configuration between $(1,1)$ and $(0,2)$:
when the electrons
are in the right dot $(0,2)$ the ground state is the spin singlet $|S(0,2)\rangle$.
The interdot tunneling coupling $\Omega$
allows coherent transitions between the
singlet spin-charge states $|S(1,1)\rangle$ and $|S(0,2)\rangle$
which are close in energy
(for $\delta=0$, the states $|S(1,1)\rangle$ and $|S(0,2)\rangle$  are in resonance).
Transitions between the triplet states $|T(1,1)\rangle$ and $|T(0,2)\rangle$
are suppressed due to large energy mismatch:
the highly energetic triplet states $|T_{\pm,0}(0,2)\rangle$  are well separated in
energy from the singlet $|S(0,2)\rangle$\cite{Taylor} ($\Delta E_{ST} \sim 400 \mu eV$)
due to tight confinement and on site exchange interaction.
In the $(1,1)$ configuration
the spin qubit subspace is formed by the singlet $|S(1,1)\rangle$ and the triplet
state $|T_{0}(1,1)\rangle$ with zero spin projection on an external magnetic field $B$;
the applied field removes the triplet degeneracy
by splitting off the $|T_{\pm}(1,1)\rangle$ states
(splitting $\Delta_{\pm} \approx 2.5 \mu\mbox{eV}$ for
the GaAs DQD described in Ref.\cite{Marcus-Sci}).
Thus, the
DQD spin system
is considered below as
a three state quantum system (qutrit).

Introducing the short notations for the relevant three states:
$|1\rangle \equiv |S(0,2)\rangle$, $|2\rangle \equiv |S(1,1)\rangle$,
$|3\rangle \equiv |T_0(1,1)\rangle$,
we write the system hamiltonian\cite{JouravlevNazarov,Taylor}:
\begin{eqnarray}
{\cal H}_{DQD}=
-\delta\ |1\rangle \langle 1|
+ \frac{\Omega}{2}\ (|1\rangle \langle 2| + |2\rangle \langle 1|)
  &&\nonumber\\
+ \varepsilon _S\ |2\rangle \langle 2| +
             \varepsilon _T\ |3\rangle \langle 3|
\label{H-2spin} .
\end{eqnarray}
The presence of the $|T_{\pm,0}(0,2)\rangle$ states induces
small exchange energies $\varepsilon _S$, $\varepsilon _T$
which can be taken into account perturbatively;\cite{Taylor}
in what follows we will neglect these
energies
that are of the order of $\Omega^2/\Delta E_{ST} \ll \Omega$.
Other possible terms in the Hamiltonian ${\cal H}_{DQD}$
(that are present for a general qutrit system\cite{CavesMilburn})
are suppressed due to conservation of spin.

In the total Hamiltonian of system plus detector
\begin{equation}
  {\cal H} = {\cal H}_{DQD} + {\cal H}_{det} + {\cal H}_{int} ,
\end{equation}
the (low transparency) QPC detector \cite{Gurvitz}
is described with
\begin{eqnarray}
{\cal H}_{det}=\sum_l E_l c_l^\dagger c_l + \sum_u E_u c_u^\dagger c_u
+\sum_{l,u} (T c_u^\dagger c_l + \mbox{H.c.})  ; &&
\label{detector}
\end{eqnarray}
here the operator $c^\dagger_l$ ($c^\dagger_u$) creates an
electron in the lower (upper) lead of the detector (Fig.\ref{schematic})
and the tunneling $T$ between leads is assumed energy independent.

The system--detector interaction can be written in analogy with
the qubit case; for a general qutrit measured by a hypothetical
detector that can distinguish all three states
one formally writes  the  interaction Hamiltonian
as proportional to the two [$SU(3)$] generators, that are diagonal
in the spin-charge basis:
\begin{eqnarray}
{\cal H}_{int}=\sum_{l,r} \left[ A\, (|1\rangle \langle 1|
         - |2\rangle \langle 2|) \otimes c_u^\dagger c_l \right. &&
  \nonumber\\
\left. + B\, (|1\rangle \langle 1|
         + |2\rangle \langle 2| - 2\, |3\rangle \langle 3|) \otimes
  c_u^\dagger c_l + \mbox{H.c.} \right]
\label{int-2spin} .
\end{eqnarray}
Introducing detector transmission probability $T_k$
for each system state $|k\rangle$,
the average currents  can be expressed
as $I_k =2\pi T_k \rho_l \rho_u e^2 V/\hbar,\ k=1,2,3$,
where $V$ is the detector bias voltage, and $\rho_{l}$ and $\rho_{u}$ are the
densities of states in the lower and upper lead.
In the case of QPC, which cannot distinguish between
states $|2\rangle \equiv |S(1,1)\rangle$ and
$|3\rangle \equiv |T_0(1,1)\rangle$
the tunneling amplitudes $A$ and $B$ are not independent:
$T_2=|T - A + B|^2=T_3=|T - 2 B|^2$,
since the corresponding average currents
are equal, $I_2=I_3\equiv I_{(1,1)}$.
For the state $|1\rangle \equiv |S(0,2)\rangle$ the average
current $I_{(0,2)}\neq I_{(1,1)}$  and it can be distinguished
from the $(1,1)$ states.

In what follows we consider a situation
when the internal detector
dynamics given by Eqs.(\ref{detector}), (\ref{int-2spin})
is much faster than the two-spin dynamics due to ${\cal H}_{DQD}$.
Also, it is assumed that the voltage applied to detector is large
so that a typical detector decoherence time
is much smaller than the typical electron tunneling times;
thus, $\hbar/eV \ll e/I_{(0,2)},e/I_{(1,1)}\ll 1/\Omega$.
The first inequality implies that coherences between
different electron passages in the QPC can be neglected.
Thus,  the  QPC detector will behave essentially classically on the typical time
scale of the DQD two-spin dynamics.

\subsection{Continuous quantum evolution according to the result}

The quantum state evolution of an open quantum system\cite{Carmichael,Davies}
has an analog in a classical state estimation procedure;\cite{Holevo}
it takes into account
the actual measurement record that is imperfectly
correlated with the system state.
In a situation when the QPC is a weakly responding detector,
$|\Delta I|\equiv |I_{(0,2)} - I_{(1,1)}| \ll I_{(0,2)}, I_{(1,1)}$, when every
tunneling electron
brings a little information about the system state,
it is reasonable to condition the system evolution on
the quasicontinuous noisy detector current $I(t)$.\cite{Kor-99-01}
Typical measurement time (that reflects accumulation of
a signal-to-noise  ratio of order one)
can be introduced\cite{Shnirman}${}^{,}$\cite{Kor-99-01},
$\tau_{meas} = 2 S_0/(\Delta I)^2 \sim 1/\Gamma$,
where $S_0\simeq 2e I_{(1,1)}$
is the low-frequency spectral density of the detector shot noise.
[We have neglected small differences
between shot noises.]
The finite measurement time implies that
given the noisy current $I(t)$, the system state is updated gradually.
Below we sketch the derivation of the measurement evolution equations
for a qutrit related to the ``informational''  Bayesian
approach.\cite{Kor-99-01,Kor-ent}

For the measurement evolution alone, the most straightforward way
is to consider the elementary act of scattering of an incoming
electron, $|i\rangle$, off the QPC tunnel barrier that depends on the charge
state of the system, $|k\rangle$.
The scattered state is expressed as a linear transform of the initial state:
$|f\rangle=\hat{S}_k |i\rangle$ where  the scattering matrix is
(see, e.g., Ref.\onlinecite{AverinSukhorukov})
\begin{equation}
\hat{S}_k=\left(
\begin{array}{cc}
  r_k  &  t_k^* \\
  t_k  & -r_k^* \\
\end{array}
\right)  ,
\end{equation}
and $r_k$, $t_k$ are the reflection and transmission amplitudes.
Following Jordan and Korotkov,\cite{JordanKorotkov}
every tunneling electron can be mapped to an ancilla qubit, whose basis states
are the scattering states: $|R\rangle$ for a reflected
electron, and $|T\rangle$
for a transmitted
electron. Scattering will entangle the states of the ancilla
qubit and the qutrit.
Then projective measurement on the ancilla will lead to %a
POVM (positive operator-valued measure) measurement
operators\cite{NielsenChuang}
of the form: $\hat{M}_R=\mbox{diag}\{r_1,r_2,r_3\}$ and $\hat{M}_T=\mbox{diag}\{t_1,t_2,t_3\}$
that satisfy the completeness condition
$\hat{M}_R^\dagger\,\hat{M}_R + \hat{M}_T^\dagger\,\hat{M}_T=1$.
The diagonal form of these operators in the qutrit basis (the spin-charge basis) follows
from the diagonal form of the interaction Hamiltonian ${\cal H}_{int}$, Eq.(\ref{int-2spin}).
For collecting electrons in the upper lead, the count of an electron
updates the qutrit density matrix according to a POVM formula
$\hat{\rho}^{'}=\hat{M}_T \hat{\rho} \hat{M}_T^\dagger/{\cal P}_T$
where ${\cal P}_T = \mbox{Tr}[\hat{M}_T^\dagger\,\hat{M}_T\hat{\rho}]=
\sum_k \rho_{kk} T_k$ is the total probability to find an
electron in the
upper lead and $T_k=|t_k|^2$ are the transmission probabilities
introduced in the previous section.
If an electron was not counted
a similar update takes place using the measurement operator $\hat{M}_R$
and the reflection probabilities, $1-T_k=|r_k|^2$.
In the basis where $\hat{M}_{R,T}$ are diagonal
these evolutions take the form of
Bayesian updates.\cite{JordanKorotkov}

The evolution rate of the system density matrix
will be related to the rate of tunneling through the QPC.
Introducing an average number $A\,\delta t$ of
(independent) tunneling attempts
per time $\delta t$,
the average current\cite{Landauer} (given the system state $|k\rangle$) is then $I_k=e A T_k$.
For times of the order of individual tunneling times, $\delta t \sim e/I_k$,
we can consider a negative-result evolution of the qutrit density matrix.
For the evolution of the diagonal elements
in case of
no tunneling we write:
\begin{equation}
\rho_{kk}(\delta t) = \rho_{kk}(0)\,(1-T_k)^{A\,\delta t}/{Norm}  .
\label{neg-rez-elementary}
\end{equation}
($Norm$ is a proper normalization.)
For a low transparency QPC, $T_k \ll 1$, the number of attempts is large,
$A\,\delta t \sim 1/T_k \gg 1$ so that
$(1-T_k)^{A\,\delta t}\approx e^{-T_k A\,\delta t}=e^{-\frac{I_k}{e}\delta t}$.
Thus, we obtain a negative-result evolution similar to Eq.(\ref{atom-Bayes1})
where the spontaneous decay rate $\Gamma_{sp}$ is replaced by the
tunneling rates $\frac{I_k}{e}$, each for every $k=1,2,3$.
The evolution of the non-diagonal elements can be shown to be of the form
similar to Eq.(\ref{atom-Bayes2}).

The above evolution can be hardly observed, since for
typical system times,
$1/\Omega,1/\Gamma \gg e/I_k$,
many electrons pass through QPC.
For $N$ independent
attempts one can derive
the conditional probability for $m$ successful tunnelings\cite{JordanKorotkov}
(per time $\tau$)
given the system is in state $|k\rangle$
\begin{equation}
{\cal P}(m,N|k) = C_N^m T_k^m (1-T_k)^{N-m}
\label{independent-probab}  ,
\end{equation}
so that the Bayesian  update of the system (qutrit)
density matrix can be shown to be of the form:
$\rho_{kk}^{'} = \rho_{kk} {\cal P}(m,N|k) /Norm$
and
$\rho_{kl}^{'} = \rho_{kl} \sqrt{\rho_{kk}^{'}\rho_{ll}^{'}/\rho_{kk}\rho_{ll}}$.

We mention that the update related to Eq.(\ref{independent-probab}),
is just a simple composition of $N$ elementary
updates, each corresponding to tunneling $(1)$ or no tunneling $(0)$ of a single electron:
e.g., for a result $\tilde{Q}=(0,1,\ldots,0)$ the measurement operator $\hat{M}_{\tilde{Q}}$
is just a multiplication of $N$ diagonal operators,
$\hat{M}_{\tilde{Q}}=\hat{M}_{R_1} \hat{M}_{T_2} \ldots \hat{M}_{R_N}$.
Also note that
Eq.(\ref{independent-probab})
sums up over $C_N^m$ identical possibilities
since QPC cannot distinguish different sequences $\tilde{Q}_1,\tilde{Q}_2,\ldots$
with the same total charge tunneled to the right lead.
By representing the (random) number $m$ of tunneling electrons per
time interval $\tau$ through the QPC current,
$m\equiv \tau \bar{I}(t,\tau)=\int_t^{t+\tau} I(t') dt'$,
and using De Moivre-Laplace limit theorem ($N\gg 1$, $T_k$ fixed)
we can replace ${\cal P}(m,N|k)$ by a Gaussian distribution for
the measurement result $\bar{I}(t,\tau)$, given the state $|k\rangle$:
$P_k(\bar{I})=\sqrt{\tau/\pi S_0} \exp{[-(\bar{I}(t,\tau) - I_k)^2 \tau/S_0]}$,
where $I_k\simeq e M T_k/\tau$ is the average current, and $S_0$
is the shot noise spectral density for a
low transparency and weakly responding QPC.
The update of the system density matrix can be written
again as a Bayesian (informational) evolution:\cite{Kor-99-01}
\begin{equation}
\rho_{kl}(t+\tau) =
\rho_{kl}(t) \frac{\sqrt{P_k(\bar{I})}\,\sqrt{P_l(\bar{I})}}{P(\bar{I})}
\label{quantBayes},
\end{equation}
with the total probability of a particular
result $\bar{I}$ given by $P(\bar{I})=\sum_k \rho_{kk}(t) P_k(\bar{I})$.

Differentiating Eq.(\ref{quantBayes}) over $\tau$ (at $\tau \to 0$)
one  obtains a stochastic equation for the qutrit state evolution
(in Stratonovich form\cite{Oksendal}).
Taking into consideration the (internal)
Hamiltonian evolution
and possible dephasing,
one can write:
\begin{eqnarray}
&&\dot{\rho}_{kl} =
\frac{\rho_{kl}}{S_0}  \sum_j \rho_{jj}\,
\left\{ (I_k - I_j) \left(I(t) - \frac{I_k + I_j}{2}\right)\right.
\nonumber \\
&&\left. \qquad\ {} + (I_l - I_j) \left(I(t) - \frac{I_l + I_j}{2}\right) \right\}
\nonumber \\
&&\qquad\ {} -\frac{\imat}{\hbar} [{\cal H}_{DQD}, \rho]_{kl}
-\gamma_{kl}\, \rho_{kl} .
        \label{Bayes1}
\end{eqnarray}
Here $I(t)$ is the formal limit (at $\tau \to 0$)
of the observed detector signal $\bar{I}(t,\tau)$.
For numerical simulations of a measurement
one complements\cite{Kor-99-01} Eq.(\ref{Bayes1}) by
\begin{equation}
I(t) = \sum_k \rho_{kk}(t) I_k + \xi(t)
\label{current-model}
\end{equation}
that is consistent with the statistics of $P(\bar{I})$;
here $\xi(t)$ is a white noise with a spectral density $S_{\xi}=S_0$.
Eq.(\ref{Bayes1}) is of the same form as an analogous
evolution equation for a system of $N$ qubits;\cite{Kor-ent}
here the summation is over the three qutrit states.
The dephasing rates, $\gamma_{kl}$,
are related to the detector non-ideality;\cite{Kor-99-01,back-action}
experimentally\cite{Buks} and theoretically\cite{Kor-99-01,back-action,AverinSukhorukov}
a QPC  is close to an ideal detector.
An SET is usually highly non-ideal\cite{Kor-99-01},
however it may reach ideality close to 1 in the co-tunneling or
Cooper pair tunneling regime.\cite{cotun}

It is worthwhile to note that
the measurement evolution
and the Hamiltonian evolution
enter in Eq.(\ref{Bayes1}) independently; they just reflect
the measurement (POVM) and unitary postulates
applied to the system at a coarse grained time $t \gg e/I_k \gg \hbar/eV$
where the evolution is noisy and quasicontinuous.
In Sec. III we will show via numerical solutions of Eq.(\ref{Bayes1})
that the continuous measurement evolution
and the Hamiltonian evolution interplay  non-trivially so that
new effective (negative-result) measurement and
Hamiltonian evolutions of the system arise at a larger time scale.

\subsection{Ensemble averaged evolution of the system}

A total ignorance of a particular measurement result $\bar{I}(t)$
corresponds to a situation when the QPC detector is considered
just as a part of a (Markovian) environment surrounding the system.
Correspondingly, the density matrix available to such
an observer (denoted as $\langle\rho_{kl}\rangle$) will be
quite different from that described by Eq.(\ref{Bayes1}).
The density matrix $\langle\rho(t)\rangle$ can be related to
$\rho(t)$, Eq.(\ref{Bayes1}) by a formal procedure of an ensemble averaging
over possible results $\bar{I}(t)$ at every time moment $t$,
similar to the classical probabilities.\cite{Holevo}
The averaging can be performed
(for a sufficiently small
time interval $\tau$), e.g., by using
the total probability of a particular result $\bar{I}$,
given by $P(\bar{I})$
and then simply adding the Hamiltonian evolution.
The result will be a standard master equation\cite{Leggett}
\begin{equation}
\langle\dot{\rho}_{kl}\rangle = -\Gamma_{kl} \, \langle\rho_{kl}\rangle
-\frac{\imat}{\hbar} [{\cal H}_{DQD}, \langle\rho\rangle]_{kl}
\label{Gen-master}
\end{equation}
with ensemble-averaged dephasing rates
\begin{equation}
\Gamma_{kl} = (I_k - I_l)^2/4 S_0 + \gamma_{kl} .
\label{decohrates}
\end{equation}
For a quantum limited (ideal) detector, $\gamma_{kl}=0$, the
dephasing rates produced due to averaging are just $(I_k - I_l)^2/4 S_0$
which is the minimum
allowed by quantum mechanics.\cite{Kor-99-01,Kor-osc,back-action,Ruskov-Bell}

The individual dephasings, $\gamma_{kl}\neq 0$,
may be a consequence of partial ignorance of the measurement
result\cite{Kor-99-01} and are parameterized by the
detector ideality (efficiency)
$\eta$ ($0\leq \eta \leq 1$):
$\gamma_{kl} = (\eta^{-1} -1)\,(I_k - I_l)^2/4 S_0$,
i.e., $\Gamma_{kl}=(I_k - I_l)^2/4 S_0\eta$.
Other sources of decoherence of the DQD system will
be discussed in Sec. IV.

\section{Emergence of negative-result evolution in the qubit subspace}

When the unitary evolution of the three-state system is taken into account
one has to explore the full Bayesian Eq.(\ref{Bayes1}),
which, as a rule, does not provide simple
solutions.
In what follows, using Eq.(\ref{Bayes1}), we will perform numerical simulation
of the measurement process for various regimes of the system-detector dynamics.
The non-trivial interplay of the quantum dynamics can be seen
if one compares the effects of measurement evolution alone vs. total evolution.
%the fact that
The measurement alone tends to collapse the system (qutrit) in either
the state $|1\rangle$ or to the qubit subspace $\{|2\rangle,|3\rangle\}$
leaving states $|2\rangle$ and $|3\rangle$ unresolved.
However, adding the
continuous coherent mixing of $|1\rangle$ and $|2\rangle$ by
${\cal H}_{DQD}$ leads to effective resolution of the states $|2\rangle$
and $|3\rangle$, so that a continuous collapse happens
either to the state $|3\rangle$ or to the remaining spin-singlet subspace.

If the continuous collapse happens to the singlet subspace,
at small
coupling $\Gamma$ the system will perform quantum oscillations,
that are weakly perturbed by the measurement.
When the coupling becomes large the picture is qualitatively different:
one reaches the regime of Zeno stabilization of the system's singlet states.
The latter is characterized by long time intervals when the QPC  current
is either $I_{(1,1)}$ or $I_{(0,2)}$, interupted by rare switching between them;
the qutrit state is correspondingly $|2\rangle \equiv |S(1,1)\rangle$
or $|1\rangle \equiv |S(0,2)\rangle$.
If the system was initially into the qubit subspace,
the continuous collapse takes the form of a slow
negative-result evolution of the qubit state
until it eventually switches to $|S(0,2)\rangle$.

The negative-result evolution is a Bayesian evolution
conditioned by the information that
the ``qubit did not switch''.
This evolution emerges in the Zeno regime as a solution
of the underlying Bayesian stochastic evolution, Eq.(\ref{Bayes1}).
It is important to note that the key feature for establishing
a negative-result evolution is
the irreversibility of the switching event
(compare with Ref.\onlinecite{Pryadko}).
Within the %Bayesian
stochastic measurement evolution, the
irreversibility is consistent with an important property
of Eq.(\ref{Bayes1}) that is seen in numerical simulations:
given a measurement record $I(t)$, two evolutions
that start from different
initial states, $\rho^{(1)}(0)$ or $\rho^{(2)}(0)$,
will become undistinguishable\cite{Kor-99-01,Ruskov-Zeno}
at a time scale of the order or greater than $1/\Gamma$.
In other words the system ``forgets''
its initial state, $\rho(0)$ so that further evolution is
dominated by the result, $I(t)$ itself.
The regime of current stabilization,
(that is essentially classical\cite{Ruskov-Bell})
then would correspond to irreversibility of the switching event.

\subsection{Collapse scenarios}

For a numerical simulation of the measurement process
we use Eq.(\ref{Bayes1}),  supplemented with the relation
for the current signal, Eq.(\ref{current-model}),
where we incorporated the current level degeneracy for
the states $|2\rangle$ and $|3\rangle$ ($I_2=I_3=I_{(1,1)}$).
In order to understand the collapse scenarios
it is instructive to look at the
stochastic equation for the relevant density matrix elements
transformed from Eq.\,(\ref{Bayes1}) to their It\^{o} form\cite{Oksendal}
(below we use $\hbar=1$).
\begin{eqnarray}
&& \dot{\rho}_{11} =
-\Omega\, \mbox{Im}\rho_{12} + \rho_{11}\,(1-\rho_{11})\,\frac{\Delta I}{2S_0}\,\xi(t), \ \
    \label{r11-Ito}\\
&& \dot{\rho}_{33} =
- \rho_{11} \rho_{33}\,\frac{2 \Delta I}{S_0}\,\xi(t),
\label{r33}
\end{eqnarray}
$\qquad\qquad\qquad\dot{\rho}_{22}=-\dot{\rho}_{11}-\dot{\rho}_{33}$, and
\begin{eqnarray}
&& \dot{\rho}_{12} =
i\, \delta\, \rho_{12} + i\, \frac{\Omega}{2}\,(\rho_{11}-\rho_{22}) -\Gamma_{12}\, \rho_{12}
\nonumber\\
&& \qquad\ \ {}  - (2 \rho_{11} - 1)\,\rho_{12}\,\frac{\Delta I}{S_0}\,\xi(t),
\label{r12-Ito}\\
&& \dot{\rho}_{13} =
i\, \delta\, \rho_{13} - i\, \frac{\Omega}{2}\,\rho_{23} -\Gamma_{13}\, \rho_{13}
\nonumber\\
&& \qquad\ \ {}  - (2 \rho_{11} - 1)\,\rho_{13}\,\frac{\Delta I}{S_0}\,\xi(t),
\label{r13-Ito}\\
&& \dot{\rho}_{23} =
 - i\, \frac{\Omega}{2}\,\rho_{13} - \rho_{11}\,\rho_{23}\,\frac{2 \Delta I}{S_0}\,\xi(t).
\label{r23-Ito}
\end{eqnarray}
Here $\Delta I \equiv I_{(0,2)}-I_{(1,1)}$ is the current difference
between the charge subspaces that can be distinguished by the QPC.
The system-detector coupling explicitly enters in Eqs.(\ref{r12-Ito}),(\ref{r13-Ito})
as $\Gamma_{12}=\Gamma_{13}\equiv (\Delta I)^2/4 S_0 + \gamma_{12}=\Gamma=(\Delta I)^2/4 S_0 \eta$

In the absence of the Hamiltonian, ${\cal H}_{DQD}=0$, the equation for $\rho_{11}$
becomes a pure noise and decouples; it can be heuristically considered as a
(position dependent) Wiener process
for the restricted variable $0\leq\rho_{11}\leq 1$ with
a diffusion coefficient, $\sim \rho_{11} (1-\rho_{11})$, approaching
a minimum (zero) at the endpoints, $\rho_{11}=0$ or $\rho_{11}=1$.
Just as in the single qubit case\cite{KorJordan-undo}
it suggests that the endpoints are the two possible attractors for
a given realization of the measurement.
In the qutrit case, the measurement also preserves the ratio
$\rho_{22}/\rho_{33}\equiv\alpha$ and the measurement evolution
happens on
a ray $\rho_{33}=(1-\rho_{11})/(1+\alpha)$ in the
physical triangle of the plane $(\rho_{11},\rho_{33})$
[$0\leq\rho_{kk}\leq1, k=1,2,3$].
Switching on the Hamiltonian mixes the density matrix components leaving
$\rho_{33}$ unaffected; geometrically this is represented as a
horizontal moving in the plane $(\rho_{11},\rho_{33})$ that
changes $\alpha$ [causing the system to change from one ray to another].
From this point of view it is easy to imagine how the new attractors
become $\rho_{33}=0$ (a horizontal ray) or the point $\rho_{33}=1$.
Finally, it is clear from Eq.(\ref{r33}) that
if the system is in either of the subspaces,
$\{|1\rangle,|2\rangle\}$ or $|3\rangle$, it will remain there:
neither Hamiltonian nor measurement evolution mixes the subspaces,
which is true for any DQD parameters, $\delta$, $\Omega$.
The decoupling of subspaces is a consequence
of the spin blockade and  indistinguishability of the states $|2\rangle$
and $|3\rangle$ by measurement.

The decoupling implies that the system will continuously
collapse to one of the subspaces under a weak continuous measurement
as seen in numerical simulations, Fig.\ref{weak}.
Since the corresponding ensemble averaged equation for $\rho_{33}(t)$
is simply $d \langle\rho_{33}(t)\rangle /dt =0$
(ensemble averaging implies just nullifying of the noise
in It\^{o} form of the equations\cite{Oksendal}),
then the ensemble averaged density matrix element is conserved:
$\langle\rho_{33}(t)\rangle  = \rho_{33}(0)$.
Thus, it must be that the fraction of members of the ensemble
that collapses to $\rho_{33}(t)=1$ (after a sufficient measurement time $t$)
is just $\rho_{33}(0)$,
which means that the
probability of collapse to either of the subspaces is $\rho_{33}(0)$
and $1-\rho_{33}(0)$  respectively (i.e., the usual probability rules apply).

\subsection{System evolution at small  coupling.
Quantum oscillations and spin blockade}
\label{small_coupling}

In order to characterize the emergence of a negative result evolution
from continuous noisy measurement
we will first consider the case when the system may establish
quantum oscillations weakly perturbed by the measurement, i.e.,
when negative-result evolution is not yet established.
In particular, this happens for small system-detector coupling,
$\Gamma \ll \Omega$ while the quantum dots detuning
$\delta$ is also small.
For simplicity, we first consider the case of an ideal detector, $\eta=1$.

\begin{figure}
\vspace*{0.1cm}
\centering
\includegraphics[width=3.2in]{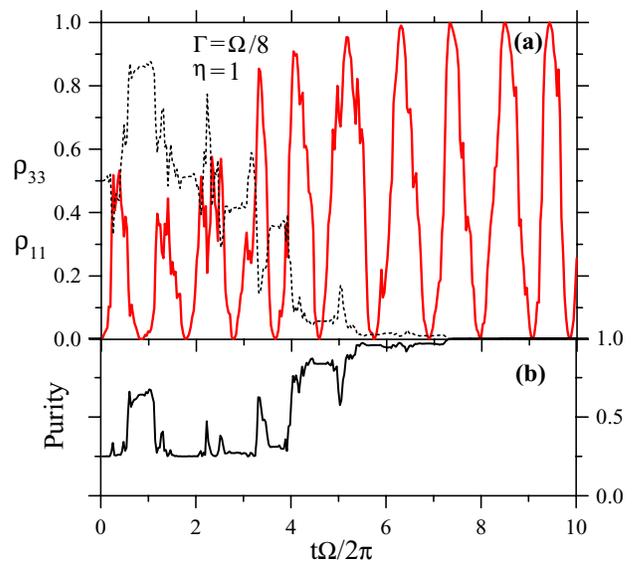}
\vspace{-0.06cm}
\caption{(Color online) Weak coupling regime, $\Gamma\ll\Omega$, at zero detuning.
(a) %A particular realization of $\rho_{33}(t)$ evolution (dashed line)
A realization of $\rho_{33}(t)$ evolution (mixed initial state)
showing collapse to the
spin-singlet subspace $\{|1\rangle,|2\rangle\}$ (dashed line);
weakly perturbed coherent oscillations of $\rho_{11}(t)$
(thick line) into the subspace.
(b) Qutrit purification due to measurement with an ideal detector, $\eta=1$.}
\label{weak}
\end{figure}

An example of
the three-state system evolution is shown
on Fig.\ref{weak}a.
In this particular realization of the measurement process,
while starting from a mixed initial state
($\rho_{11}(0)=0$, $\rho_{22}(0)=\rho_{33}(0)=1/2$,
$\rho_{ij}(0)=0, i\neq j$)
$\rho_{33}(t) \to 0$ (dashed line on Fig.\ref{weak}a),
i.e., the system is continuously collapsed
to the spin-singlet subspace,
$\{|1\rangle\equiv|S(0,2)\rangle,|2\rangle\equiv|S(1,1)\rangle\}$
after a transition time of the order of $1/\Gamma$.
On the same time scale,
weakly perturbed quantum
oscillations are established in this subspace
[solid line on Fig.\ref{weak}a) for $\rho_{11}(t)$].

The oscillation scenario in case of collapse to the singlet
subspace is easily understood.
Once $\rho_{33}(t)$  gets to zero,
Eqs.(\ref{r11-Ito}),(\ref{r12-Ito}) just become
of the same form as that describing
quantum oscillations for a single qubit.\cite{Kor-osc}
For zero detunning, $\mbox{Re}\rho_{12}$ decouples ($\mbox{Re}\rho_{12}(t)\rightarrow 0$),
while $\rho_{11}(t)-\rho_{22}(t)=\cos[ \Omega t +\varphi(t)]$
and $2 \mbox{Im}\rho_{12}(t)=\sin[ \Omega t +\varphi(t)]$ oscillate
with a unity amplitude (for $\eta=1$), with phase $\varphi(t)$
that slowly diffuses in time.\cite{Kor-osc}
Correspondingly, the oscillating scenario can be distinguished
by the average current $\langle I(t) \rangle_t = (I_{(0,2)} + I_{(1,1)})/2$
or by the current spectral density
$S_I(\omega) = 2\int_{-\infty}^{\infty} K_I(\tau)\exp(i\omega \tau) d\tau$
(with $K_I(\tau)=\langle I(t) I(t+\tau)\rangle_t - \langle I(t) \rangle_t^2$ being the
current correlation function).
Since the evolution of $\rho_{33}(t)$ is transitional
in character (see Fig.\ref{weak}),
it will not affect the long-time average $K_I(\tau)$.
Then using, e.g., the methods of Ref.\onlinecite{Ruskov-Zeno},
it can be shown that the detector power spectrum will have
the same form as in the one-qubit case\cite{Kor-osc}
for any DQD parameters, $\delta$, $\Omega$.
In particular, in the weak coupling regime (for $\delta=0$)
the spectrum $S_I(\omega)$ exhibits a Lorenzian
peak at the Rabi frequency $\Omega$
with a signal-to-noise ratio of $4\eta$,
and a width $\Gamma = (I_{(0,2)} - I_{(1,1)})^2/4S_0\eta$.

On (Fig.\ref{weak}b) the qutrit purity is plotted.
[We have defined it
as $\mbox{Pur}(t) = (3\mbox{Tr}[\hat{\rho}(t)^2] - 1)/2$, so that $\mbox{Pur}=1$
for a pure state and $\mbox{Pur}=0$ for a totally mixed state.]
It is seen that the qutrit is eventually
reaching a pure state [even though with a random phase, $\varphi(t)$] for a time
of the order of $1/\Gamma$.
The purification of the qutrit state
is yet another demonstration of non-trivial interplay of dynamics.
Indeed, Hamiltonian evolution alone conserves purity,
while a measurement alone leaves states $|2\rangle$ and $|3\rangle$ unresolved.
The purification of the state is due to
an effective resolution of the states $|2\rangle$
and $|3\rangle$ and the fact that
no information is lost with an ideal measurement.

For a measurement with a non-ideal detector, $\eta < 1$,
simulations show (if collapse happened to the spin-singlet subspace)
that the amplitude of quantum oscillations is less than $1$ and fluctuates
in time. Correspondingly,
the average purity $\langle \mbox{Pur}(t)\rangle_t$
saturates at some lower value.
Using Eqs.(\ref{r11-Ito})-(\ref{r23-Ito}) we have derived
in the weak coupling regime
(compare with Ref.\cite{QinRusKor})
$\langle Pur(t)\rangle_t \simeq 1 +(3/4)[1/2\eta-\sqrt{(1+1/2\eta)^2-2}]$
so that the qutrit state remains mixed;
for small $\eta$ the qutrit purity approaches $1/4+3/2\eta$.
If collapse happens to $\rho_{33}(t) \to 1$,
the state will purify even for measurement with a non-ideal detector.
In this scenario the average current is $I_{(1,1)}$ and the power spectrum
is flat, $S_I(\omega)=S_0$
[the detector signal is just $I(t) = I_{(1,1)} + \xi(t)$].

\subsection{Large coupling. Zeno stabilization
and emergence of negative-result evolution}

\begin{figure}
\vspace*{0.1cm}
\centering
\includegraphics[width=3.2in]{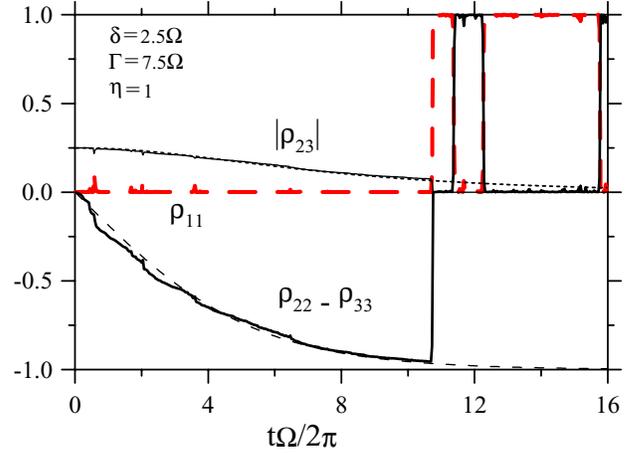}
\vspace{-0.06cm}
\caption{(Color online) Strong coupling regime, $\Gamma\gg\Omega$.
Zeno stabilization in the qubit subspace, $\rho_{11}=0$,
for a long time until it switches to $\rho_{11}=1$ (thick dashed line).
The noisy qubit evolution of a mixed initial state
%($\rho_{22}(0)=\rho_{33}(0)=0.5$, $\mbox{Im}\rho{23}(0)=0.25$)
shown for $\rho_{22}-\rho_{33}$ and $|\rho_{23}|$
vs. postulated negative-result evolution given by Eqs.(\ref{SW-rate})-(\ref{Bayes-upd2})
(long dashed and short dashed lines).
}
\label{strong-Zeno}
\end{figure}

     We now turn to the case of relatively large system detector
coupling, $\Gamma \gg \Omega$; the DQD detuning energy, $\delta$, can be
either small or large.
Numerical simulations confirm that collapse scenarios
remain the same in the large coupling regime
consistent with our argumentation
in Sec. III A.
However, in the strong coupling case the evolution qualitatively changes;
the typical collapse time to
either of the subspaces, \{$|1\rangle,|2\rangle$\} or $|3\rangle$,
becomes much longer than $1/\Gamma$.
Also, instead of quantum oscillations within the
spin-singlet subspace $\{|1\rangle\equiv|S(0,2)\rangle$, $|2\rangle\equiv|S(1,1)\rangle\}$,
associated with $\Omega$,
we have relatively long stabilization of the system state (Fig. \ref{strong-Zeno})
in one of the two states since the measurement is trying to localize
and ``freeze'' the system in a definite charge state.
This is a manifestation of the quantum Zeno effect:\cite{Khalfin,MisraSudarshan}
in the case of continuous measurement the detector is always coupled
to the system, so the approach to quantum Zeno regime corresponds
to the limit of stronger and stronger coupling.
For a finite  coupling
the long stabilization periods will be interrupted by rare switching events
between the subspaces (the switching transition
time is of the order of $1/\Gamma$; see below).

Anticipating the scenario of a negative-result evolution in
the qubit subspace outlined in the beginning of Sec. III,
we turn to the calculation of the switching rate
between the two singlet states,
which will set the time scale of non-unitary dynamics.
From an ensemble averaged point of view the qubit ``decay''
from $|S(1,1)\rangle$ to $|S(0,2)\rangle$
(triplet state $|T_0(1,1)\rangle$ is spin blocked)
is described as a smooth decrease of $\langle\rho_{11}(t)\rangle$
and therefore the switching rate would
be most easily obtained if we consider the master equation (\ref{Gen-master})
that follows from averaging of measurement evolution.

Due to current level degeneracy,
$I_2=I_3=I_{(1,1)}$,
the dephasing rate $\Gamma_{23}$, Eq.(\ref{decohrates}),
vanishes just because the detector cannot distinguish between
the corresponding states
(no information can be obtained and therefore no information can be lost).
Introducing     %$\Delta I \equiv I_1 - I_{23}$
$\Gamma_{12}=\Gamma_{13}=\Gamma \equiv (\Delta I)^2/4 S_0 \eta$
the master equation reads
\begin{eqnarray}
\lefteqn{ \langle\dot{\rho}_{kl}\rangle =
-\frac{\imat}{\hbar} [{\cal H}_{DQD}, \langle\rho\rangle]_{kl}  }
&&   \label{Master}  \\
&& \qquad {} - \Gamma\ \left(
\begin{array}{ccc}
  0        & \langle\rho_{12}\rangle & \langle\rho_{13}\rangle \\
\langle\rho_{21}\rangle  & 0         & 0 \\
\langle\rho_{31}\rangle  & 0         & 0 \\
\end{array}
\right)_{kl} \nonumber
\end{eqnarray}
In order to calculate the switching rate from the state $|1\rangle =|S(0,2)\rangle$
we start with the initial condition $\rho_{11}=1$. Using Eq.(\ref{Master})
one derives the small time evolution:
\begin{equation}
\langle\rho_{11}(t)\rangle = 1 - \frac{\Omega^2 t^2}{4} + \cdots
\label{small-t} .
\end{equation}
Notice, that there is no linear term in the expansion, Eq.(\ref{small-t}),
which tells us that the decay is not exponential at a small time scale.
This fact
makes quantum Zeno effect physics possible.
The $t^2$-coefficient turns out to be
$-(\Omega)^2/4 = -\mbox{Tr}[\hat{\rho}^2 {\cal H}^2_{DQD}] +
\mbox{Tr}[\hat{\rho}{\cal H}_{DQD}\hat{\rho}{\cal H}_{DQD}]$,
i.e. it is determined by coherent (Hamiltonian) evolution alone,
consistent with the discussion
in Refs. \onlinecite{Khalfin,MisraSudarshan,Peres}.

It is instructive to find an approximate solution for $\langle\rho_{11}(t)\rangle$ considering
in (\ref{Master}) terms proportional to $\Omega$ as a small perturbation ($\Omega \ll \Gamma$,
for strong coupling).
To first order in the perturbation $\Omega$ we obtain
for $\mbox{Im}\langle\rho_{12}(t)\rangle$:
\begin{eqnarray}
\lefteqn{\mbox{Im}\langle\rho_{12}(t)\rangle =e^{-\Gamma t}\,\left[\mbox{Re}\rho_{12}(0)\,\sin{\delta t} +
\mbox{Im}\rho_{12}(0)\,\cos{\delta t} \right] }
&& \nonumber\\
&& \ {}+ \frac{\Omega}{\Gamma^2+\delta^2}\,\left\{\frac{\Gamma}{2}
+ e^{-\Gamma t}\,\left[ \delta\,\sin{\delta t}- \Gamma\cos{\delta t}\right] \right\}.
\label{pert-omeg}
\end{eqnarray}
Using the relation,
$\langle\rho_{11}(t)\rangle=\rho_{11}(0) - \Omega \int_0^t \mbox{Im}\langle\rho_{12}(t')\rangle dt'$,
we can obtain an approximate solution for $\langle\rho_{11}(t)\rangle$ and its small $t$
expansion coincides with first few terms of the expansion,
Eq.(\ref{small-t}).
However on a time scale $t^* \gtrsim 1/\Gamma$, as seen in Eq.(\ref{pert-omeg}),
the exponential terms drop out and one reaches an
expansion that has a linear term:
$\langle\rho_{11}(t^*)\rangle \simeq 1 - \frac{\Omega^2\Gamma}{2(\Gamma^2+\delta^2)}\,t^*$.
Similarly, one can show that if one starts from    an
initial state in the
\{$|2\rangle$, $|3\rangle$\}-subspace then on a coarse grained time scale
$\langle\rho_{22}(t^*)\rangle \simeq
\rho_{22}(0)\,\left[1 - \frac{\Omega^2\Gamma}{2(\Gamma^2+\delta^2)}\,t^*\right]$;
$\langle\rho_{33}(t)\rangle$
is just conserved.
By solving numerically the master equation (\ref{Master}) for
$\langle\rho_{11}\rangle$, $\langle\rho_{22}\rangle$,
$\mbox{Re}\langle\rho_{12}\rangle$, $\mbox{Im}\langle\rho_{12}\rangle$ one confirms that
the decay of the subspaces is indeed exponential for large times
and the switching rate between the subspaces is
\begin{equation}
\Gamma_{sw} =\frac{\Omega^2\Gamma}{2(\Gamma^2+\delta^2)}
\label{SW-rate} .
\end{equation}
Note that the strong coupling limit when $\Gamma \gg \Omega$ ($\delta$ is arbitrary)
implies that $\Gamma_{sw} \ll \Gamma$, i.e. the subspace life time
is much longer than both $1/\Gamma$ and $1/\Omega$ implying
quantum Zeno stabilization.\cite{EnsAverageView}
The exponential decay at times $t^* \gtrsim 1/\Gamma$ is a sign
of the irreversibility of the measurement\cite{Peres-book}
that appears as a switching event.

Having calculated the switching rate between the spin-charge states
we can write (postulate) an ansatz for the
time evolution of $\rho(t)$ according to a given result.
Starting from the qubit subspace,
for times $t^* \gtrsim 1/\Gamma$
one will be able to discriminate between the two current values
($I_{(1,1)}$ or $I_{(0,2)}$) and thus to
distinguish whether the system has decayed (switched)
to the third state or not.
The conditional probability for the state $|2\rangle = |S(1,1)\rangle$
not to decay by time $t$ is given by $P_2(t)=\exp{(-\Gamma_{sw}t)}$,
while analogous probability for $|3\rangle = |T_0(1,1)\rangle$ is
$P_3(t)=1$ (due to spin-blockade).
Using the quantum Bayes  rule,
similar to the two-level atom, Eqs.(\ref{atom-Bayes1}),(\ref{atom-Bayes2})
[see also Eq.(\ref{quantBayes})],
one can write  the effective negative-result evolution of
the spin qubit subsystem
given that it did not decay by time $t$:
\begin{eqnarray}
&&\rho_{22}( t) = \frac{\rho_{22}(0)\, P_2(t)}{P_{tot}(t)} ,\
\rho_{33}( t) = \frac{\rho_{33}(0)}{P_{tot}(t)} \ \
    \label{Bayes-upd1}\\
&& \rho_{23}( t) = \rho_{23}(0)\,
\sqrt{\frac{\rho_{22}(t)\,\rho_{33}(t)}{\rho_{22}(0)\,\rho_{33}(0)}}
\ e^{ -i \phi(t,\delta,\Omega,\Gamma) } ,  \qquad
\label{Bayes-upd2}
\end{eqnarray}
where the total probability not to decay is given by
$P_{tot}(t)=\rho_{22}(0)\, P_2(t) + \rho_{33}(0)\, P_3(t)$,
and $\phi(t,\delta,\Omega,\Gamma)$ is an accumulated phase (see below).

Consistency of the informational (Bayesian) approach requires
that the negative-result ansatz for the $\rho(t)$-evolution be
reproduced by the underlying evolution, Eq.(\ref{Bayes1}),
according to the noisy record $I(t)$.
In Fig.\ref{strong-Zeno} we show the negative-result evolution
for $\rho_{22}(t) - \rho_{33}(t)$ and $|\rho_{23}(t)|$
defined by Eqs.(\ref{SW-rate}),(\ref{Bayes-upd1}),(\ref{Bayes-upd2}),
versus the density matrix evolution generated through
simulation of the noisy measurement process via Eqs.(\ref{Bayes1}),(\ref{current-model}).
The (noisy) evolution in the qubit subspace
is quite regular, and the state eventually approaches
$|3\rangle \equiv |T_0(1,1)\rangle$ (if the system did not switch).
One can see that the two evolutions well agree in the
strong coupling regime; the agreement is already established at
$\Gamma \gtrsim 5 \Omega$.
The agreement reveals a non-trivial property of the Bayesian stochastic
evolution Eq.\,(\ref{Bayes1}).
It means that the postulated
negative-result evolution given by  Eqs.(\ref{Bayes-upd1}), (\ref{Bayes-upd2}),
can be actually derived from Eq.\,(\ref{Bayes1}) in the
Zeno regime,
as an interplay of
measurement evolution and Hamiltonian evolution at an underlying microscopic level.

The time scale of the negative-result evolution is set by the
switching rate $\Gamma_{sw}$, Eq.(\ref{SW-rate}),
in which the detector rate
$\Gamma=(\Delta I)^2/4 S_0 \eta \equiv (\Delta I)^2/4 S_0 + \gamma_{12}$
is the total rate, including the measurement rate and the additional rate
due to detector non-ideality.
Via numerical simulations we have confirmed the dependence of
$\Gamma_{sw}$ on $\eta$.
We note that while the
rate $\gamma_{12}$ would lead
to dephasing
in the spin-singlet subspace \{$|1\rangle,|2\rangle$\} [see Eq.(\ref{r12-Ito})],
it affects the spin qubit coherently\cite{non-ideality-neg}
in the sense that Eq.(\ref{Bayes-upd2}) preserves the coherence ratio
$|\rho_{23}|/\sqrt{\rho_{22}\,\rho_{33}}$.
The reason is
the indistinguishability of the states
$|S(1,1)\rangle$ and $|T_0(1,1)\rangle$ by the measurement.
Particularly, a pure state remains pure,
which we have confirmed numerically.
(A similar conclusion was drawn in Ref.\onlinecite{Pryadko} from a completely
different viewpoint.)
The mixed state will generally purify
(see the discussion in Ref.\onlinecite{Ruskov-neg}).
The final qubit (and qutrit) purification happens on
the same time scale $\Gamma_{sw}^{-1} \gg \Gamma^{-1}$
as the collapse to the spin-singlet or
spin-triplet subspaces.

The update for the non-diagonal element $\rho_{23}$ in (\ref{Bayes-upd2})
reflects not only
the conservation of coherence, but  includes an accumulated
phase $\phi(t,\delta,\Omega,\Gamma)$, which remains undefined
by the negative-result ansatz.
Stochastic numerical simulations by Eqs.(\ref{Bayes1}),(\ref{current-model})
show that the phase is linear in time
even for small detuning, $|\delta| \lesssim \Omega$:
$\phi(t) = \varepsilon^{\mbox{\small eff}}_{23}(\delta,\Omega,\Gamma)\,t$
and vanishes for $\delta=0$
(while noisy,
it stabilizes just for times $\gtrsim 1/\Gamma$).
The coefficient $\varepsilon^{\mbox{\small eff}}_{23}$ can be interpreted as
an energy splitting
between the spin qubit states $|2\rangle$ and $|3\rangle$
induced by the negative-result measurement in the presence
of the localized singlet state, $|1\rangle$.
For small detuning, $\delta \sim \Omega$,
the energy splitting
is small,
$|\varepsilon^{\mbox{\small eff}}_{23}|\ll  |\varepsilon^{pert}_{23}|\simeq  \frac{\Omega^2}{4|\delta|}$;
it has the same sign as the exchange splitting, $\varepsilon^{pert}_{23}$, that would be
induced perturbatively.

The effective energy splitting can be derived if one compares
ensemble averaging of the
negative-result evolution given by Eqs.(\ref{Bayes-upd1}),(\ref{Bayes-upd2})
with the ensemble averaged evolution of the ``original''
stochastic equations (\ref{Bayes1}).
Since the negative-result evolution
is the Bayesian evolution at a coarse grained time scale $t^* \gtrsim 1/\Gamma$,
its ensemble averaging must coincide with the
evolution given by the master equation (\ref{Master})
considered at times $\gtrsim 1/\Gamma$.
Averaging of the negative-result evolution is straightforward using the probability
not to decay, $P_{tot}(t)$. Starting from the qubit subspace it gives, e.g., for the
non-diagonal matrix element:
\begin{equation}
\langle\rho_{23}(t)\rangle = \rho_{23}(0)\, e^{-\Gamma_{sw}\,t/2}\,
e^{-i\,\varepsilon^{\mbox{\footnotesize eff}}_{23}\, t} .
\label{r23-ens-large-t}
\end{equation}
On the other hand Eq.(\ref{Master}) gives for the
ensemble averaged evolution of
$\langle\rho_{23}(t)\rangle$, $\langle\rho_{13}(t)\rangle$:
\begin{eqnarray}
&&\langle{\dot{\rho}}_{23}(t)\rangle = - i\,\frac{\Omega}{2}\langle\rho_{13}(t)\rangle
    \label{r23-ens}\\
&& \langle{\dot{\rho}}_{13}(t)\rangle = i\, \delta\,
\langle\rho_{13}(t)\rangle - i\,\frac{\Omega}{2}\,
\langle\rho_{23}(t)\rangle - \Gamma\, \langle\rho_{13}(t)\rangle
\label{r13-ens} . \qquad\
\end{eqnarray}
It can be solved exactly, and for the initial values $\rho_{13}(0)=0$,
$\rho_{23}(0)$,
we obtain:
\begin{equation}
\langle\rho_{23}(t)\rangle = \rho_{23}(0)\, e^{-\tilde{\Gamma} t/2}\,
\left( \cosh{\tilde{\Omega} t} +
\frac{2\tilde{\Gamma} }{\tilde{\Omega} } \,
 \sinh{\tilde{\Omega} t} \right)
\label{complex-sol}
\end{equation}
where $\tilde{\Gamma} \equiv \Gamma - i\,\delta$,
$\tilde{\Omega} \equiv \sqrt{\tilde{\Gamma}^2-\Omega^2}/2$.
Taking the strong coupling limit $\Gamma \gg \Omega$
($\delta$ is arbitrary), at
times $t^* \gtrsim 1/\Gamma$ when some contributions to Eq.(\ref{complex-sol})
are exponentially suppressed [similar to Eq.(\ref{pert-omeg})],
we reproduce the evolution for $\langle\rho_{23}(t)\rangle$,
Eq.(\ref{r23-ens-large-t}),
with an energy splitting:
\begin{equation}
\varepsilon^{\mbox{\small eff}}_{23}(\delta,\Omega,\Gamma)
= \frac{\Omega^2 \delta}{4 (\Gamma^2 + \delta^2)}
= \Gamma_{sw} \frac{\delta}{2 \Gamma} ;
\label{eff-split}
\end{equation}
it approaches $\varepsilon^{pert}_{23} \simeq  \frac{\Omega^2}{4\delta}$
for large $\delta$.
Eq.(\ref{eff-split}) is confirmed with a
very good accuracy by the results
obtained through direct simulation of the stochastic Bayesian
evolution equations (\ref{Bayes1}).
Eqs.(\ref{Bayes-upd1}),(\ref{Bayes-upd2}),(\ref{eff-split}) suggest that
the spin-charge states under continuous strong measurement
can be re-interpreted as the ``new'' energy states, if the quantum
evolution is monitored at times $t^* \gtrsim 1/\Gamma$.

\subsection{Zeno stabilization at small coupling and large detuning}

\begin{figure}
\vspace*{0.1cm}
\centering
\includegraphics[width=1.8in]{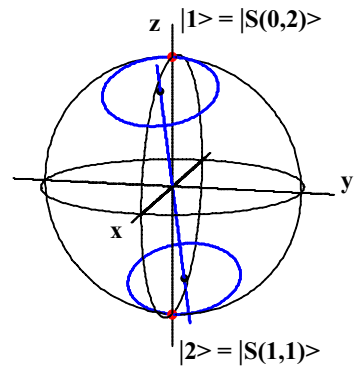}
\vspace{-0.06cm}
\caption{(Color online) Bloch sphere in the spin-singlet subspace illustrates
the Zeno stabilization %of either of the singlets
in case of weak coupling and large detuning.
%$\Gamma \ll \Omega \ll \delta$.
The projection of the Hamilton
evolution (rotation around an axis close to $z$-axis) along the
meridians (measurement evolution) is effectively small.
}
\label{kinematic-Zeno}
\end{figure}

Interestingly, Zeno stabilization can take place
even at small coupling, if the detuning is sufficiently large,
$\Gamma \ll \Omega \ll \delta$.
Qualitatively, this can be understood
by kinematical reasons as illustrated on the spin-singlet Bloch sphere,
Fig.\ref{kinematic-Zeno}. In the stochastic Bayesian equations (\ref{Bayes1}),
for large detuning $\delta \gg \Omega$,
the Hamiltonian evolution is a fast rotation
with $\Omega_{sys}=\sqrt{\delta^2 + \Omega^2}$
around an axis close to the $z$-axis.
On the other hand, the measurement alone tries to localize
the state to either of the poles (the singlet states) acting along the
meridians.
The angular velocity along the meridians is $\Omega \ll \Omega_{sys}$;
also the oscillation amplitude is $\sim \frac{\Omega}{\delta}\ll 1$,
so that the Hamiltonian evolution is effectively suppressed due to averaging.
Thus, even though the coupling is small, $\Gamma \ll \Omega$,
it might be large with respect to
an effective evolution rate along the meridians
which quantifies a quantum Zeno effect.
Numerical simulation of noisy measurement evolution shows that
the DQD two-electron system either collapses to the spin-singlet
subspace (similar to that shown in Fig.\ref{strong-Zeno}),
or ends up at the triplet state.
In the first scenario, the system is stabilized for a relatively long time
$\Gamma_{sw}^{-1} \gg \Gamma^{-1}$ in one of the singlet
states while performing rare switchings between them.

\begin{figure}
\vspace*{0.1cm}
\centering
\includegraphics[width=3.2in]{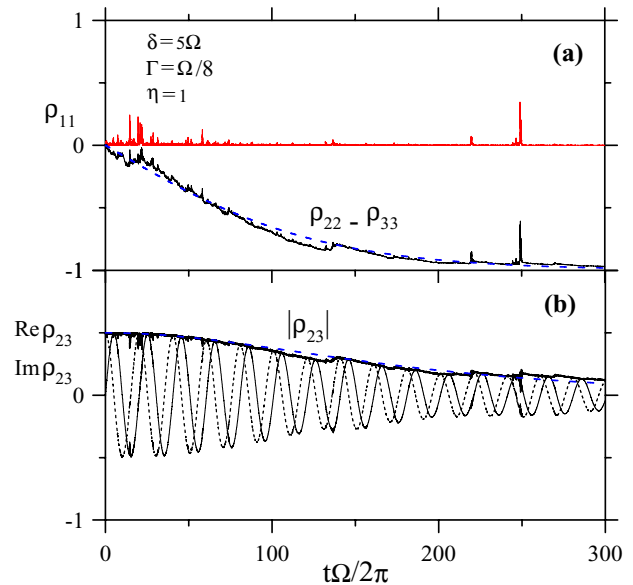}
\vspace{-0.06cm}
\caption{(Color online) Evolution in the regime of small
coupling and large detuning. %$\Gamma \ll \Omega \ll \delta$.
(a) A realization when the qubit
did not switch: $\rho_{11}(t)$ remains zero always. %(despite unresolved attempts).
Evolution of $\rho_{22}-\rho_{33}$ shows collapse
to the state $|3\rangle\equiv|T_0(1,1)\rangle$.
(b) Oscillation of coherences, $\mbox{Re}\rho_{23}$ %(thin dashed line)
and $\mbox{Im}\rho_{23}$ (thin dashed and solid lines), and their
envelop $|\rho_{23}|$ (thick solid line) vs.
negative-result  evolution of $|\rho_{23}|$ (dashed line)
according to Eqs.(\ref{SW-rate})-(\ref{Bayes-upd2}).
}
\label{large-detun}
\end{figure}

The switching rate $\Gamma_{sw}$ is given by the same formula,
Eq.(\ref{SW-rate}), since the $\Omega$-terms in
the master equation are still
small perturbations with respect to the remaining terms
as in Eq.(\ref{pert-omeg}).
Solution of the  master equation (\ref{Master}) confirms that
at a time scale $t^* \geq 1/\Gamma$ the decay from the subspaces is exponential
as was in the case of a strong coupling, with the same switching rate.
Correspondingly,
numerical simulation of noisy measurement evolution
of the DQD two-electron system, as shown on Fig.\ref{large-detun}a,b,
is in a good agreement with the negative-result evolution described
by Eqs.(\ref{Bayes-upd1}),(\ref{Bayes-upd2}).
Generally, the agreement is established already at $\delta \gtrsim 3 \Omega$.
Here, the initial qubit state was chosen to be a pure state and it remains pure
as well as the purity of the total qutrit state (not shown).
The negative-result evolution in the spin qubit subspace takes place
as long as the system did not switch to the third state.
In Fig.\ref{large-detun}
a realization
of the measurement process is shown when the system did not switch at all.
Indeed, in a situation when only a $\sigma_z$-evolution is present
in addition to the negative-result evolution,
Eqs.(\ref{Bayes-upd1}),(\ref{Bayes-upd2}),
the probability not to decay
is given by $P_{tot}(t)=\rho_{33}(0)+\rho_{22}(0) e^{-\Gamma_{sw} t}$.
So, it approaches $\rho_{33}(0)$, remaining finite for large times.

The induced energy splitting leads to a
$\sigma_z$ evolution
for $\mbox{Re}\rho_{23}$, $\mbox{Im}\rho_{23}$, whose amplitude envelope
is provided by the negative-result evolution, Fig.\ref{large-detun}b.
For large detuning and small coupling the energy splitting
is mainly due to perturbative influence of the third state:
$\varepsilon^{\mbox{\small eff}}_{23} \simeq \varepsilon^{pert}_{23} = \Omega^2/4\delta$
as it follows from Eq.(\ref{eff-split}).

\section{Available experimental parameters}

{\it 1. Coupling strength and switching rate.}
The possibility to detect single electronic charges
in quantum dots (QD) via  quantum point contact
or single electron transistor
was recently demonstrated in experiments
of different groups. %\cite{Buks}
\cite{Marcus-Sci,Kouw-prl94,Rimberg-nat}
The detector coupling,
$\Gamma = (\Delta I)^2/4 S_0\eta$, essentially determines
the measurement time
$\tau_{meas} \sim 1/\Gamma$ to reach a signal-to-noise ratio
of order one, where $\Delta I$ is the current difference
signal in the detector due to the presence (or absence) of
an extra electron charge $e$ in the QD.
For the charge sensitivity of an rf-SET
$\delta q = 2.4\times 10^{-5} e/\sqrt{Hz}$,
provided in Ref.\onlinecite{Rimberg-nat},
one estimates $\Gamma^{SET}_{exp} \approx 10^7\, s^{-1}$.
Recent experiment of Rimberg group\cite{Rimberg-new2}
demonstrated a charge sensitivity of an rf-SET of
$\delta q = 2.4\times 10^{-6} e/\sqrt{Hz}$.
Due to quadratic dependence on sensitivity,
$\Gamma \propto (\delta q)^{-2}$, this amounts to a
two orders of magnitude improvement,\cite{improvement}
$\Gamma^{SET}_{exp}\simeq 10^{9}\  s^{-1}$.
For an rf-QPC,\cite{rf-QPC} we estimated
$\Gamma^{QPC}_{exp} \simeq 10^{6}\  s^{-1}$.
The high detector ideality of a QPC also may be lost in the rf-regime.\cite{Gambetta}

The above estimates show that, in principle, both a weak coupling
as well as a strong coupling regime of a negative-result evolution
are experimentally reachable.
For a typical tunneling,\cite{Taylor} $t_c \approx 10\ \mu eV$
the characteristic frequency
$\Omega = 2\, t_c/\hbar \approx 3\times 10^{10}\ s^{-1}$ is large.
Given the range of DQD detuning, $|\delta| \approx 0 - 300\ \mu eV$,
one can reach a quantum Zeno stabilization if a relatively large detuning is taken.
Smaller values of tunneling, $t_c \approx 1\ \mu eV$ or even $t_c \approx 0.1\ \mu eV$,
are also reachable\cite{Kouwenhoven-Sci} so that a
strong coupling regime may be realized.
Typical values of switching rate $\Gamma_{sw}$ for the presented parameters are
in the range $2\times 10^{5} - 5\times 10^{7} s^{-1}$.

{\it 2. Decoherence due to charge fluctuations.}
     The decoherence in the DQD two-electron system has various sources.
The coupling to uncontrollable detector degrees of freedom leads
to detector non-ideality, $\eta$, partially discussed in  Section II.
Another mechanism of decoherence is due to systems's coupling
to background charge fluctuations\cite{CoishLoss,HuDasSarma,Taylor,RomitoGefen}
that will lead to a non-zero dephasing $\gamma_{23}$ in the spin-qubit subspace.
It was argued\cite{CoishLoss,HuDasSarma,Taylor} that such dephasing will be well suppressed
in the far-detuned regime $\delta \gg t_c$, where it may be
$\gamma_{23}\approx 10^3 - 10^5\ s^{-1}$.
However, close to the charge degeneracy, $\delta \lesssim t_c$, the dephasing
may strongly increase\cite{HuDasSarma,RomitoGefen} to
$\gamma_{23}\approx 10^6 - 10^7\ s^{-1}$ due to higher sensitivity to fluctuations
of the DQD parameters.
In the latter case it will be difficult to see the effect of Zeno stabilization
since $\gamma_{23}$ becomes comparable to the switching rate $\Gamma_{sw}$.

{\it 3. Decoherence due to phonons.}
Yet another decoherence mechanism (also related to charge degrees of freedom)
is due to coupling  to a phonon  %boson
environment.\cite{LeggettRevModPhys,Brandes}
Physically, the relevant process is the double-dot inelastic
tunneling\cite{Marcus-Sci,Kouwenhoven-Sci}
from state $|2\rangle = |S(1,1)\rangle$ to state $|1\rangle = |S(0,2)\rangle$
quantified by the inelastic rate $\Gamma_{in}(\delta)$.
Relaxation process associated with the inelastic tunneling may lead
to a contribution to the switching rate $\Gamma_{sw}$,
assuming a weak environment, $\Gamma_{in}\ll \Omega,\Gamma$.
Associated contribution to the dephasing $\gamma_{12}$
(also of the order of $\Gamma_{in}$)
is expected not to destroy the negative-result evolution
in this case.

Estimations of the inelastic rate\cite{Taylor} give the range of $0.01 - 100 neV$
(corresponding to approximately $10^{4} - 10^{8}\ s^{-1}$) depending on the
energy splitting $\Delta E$ between the relevant states.
Due to generic cubic dependence on the splitting, $\Gamma_{in} \propto (\Delta E)^3$,
one can hope to find a range of not too large detuning $\delta$ so that
$\Gamma_{in}$ is in the range of  $10^{4} - 10^{5}\ s^{-1}$.
Thus, eventually one may expect
to manage the inequality $\Gamma_{in} \ll \Gamma_{sw}$,
so that the Zeno stabilization will effectively ``fight'' against decoherence.
We note that various models of boson environment may affect
the predicted negative-result evolution in different ways
that deserve
a separate investigation.

{\it 4. Implications of the QD nuclei.}
The contact hyperfine interaction of the electron spins with the surrounding
nuclei spins  in the DQD leads to entanglement with the uncontrolable
spin-bath degrees of freedom;\cite{Dobrovitski,Dobrovitski1,Zurek,Glazman}
in quasi-classical language
it can be described as an effect of ``inhomogeneous broadening''
quantified by the random
nuclear Overhauser field.
\cite{nuclear-quasistatic,Taylor,Dobr1,Dobrovitski-rev}
The dephasing caused by such effects in the spin qubit have been
experimentally measured\cite{Marcus-Sci,HansonReview}
(dephasing time is of the order of $\sim 10^{-8}\ s$);
such strong dephasing may conceal any interesting quantum evolution.
For a non-interacting bath  it was shown that
this type of decoherence
can be completely eliminated by using various spin-echo techniques.\cite{Witzel}
Realistically, for a weakly interacting bath, application of such techniques
would reduce the decoherence by a few orders of magnitude.\cite{Witzel,Zhang}
For a GaAs DQD spin qubit system a ``true'' decoherence
time of the order of $\sim 10^{-6}\ s$ was measured experimentally
using simple  spin-echo.\cite{Marcus-Sci,HansonReview}

This suggests that spin-echo techniques could be of use
also to reveal the negative-result evolution
we are discussing in this paper.
We leave for a future project
the investigation of various possibilities for application of
spin-echo techniques in conjunction with a negative-result measurement
evolution in the context of manipulation and/or preparation
of the state of a DQD spin qubit.

\section{Conclusion}

We have shown that a negative-result evolution
of a spin qubit
can effectively emerge out of noisy measurement of a DQD spin system
by a linear detector such as a quantum point contact. The evolution
emerges as an interplay of measurement (non-unitary) dynamics
and Hamiltonian dynamics of the three-level system,
when quantum Zeno stabilization of the spin qubit subspace takes place.

Besides implications to the theory of quantum measurements,
our results may be of practical use for manipulation of a DQD spin qubit
(for papers discussing
quantum measurements as an important resource see, e.g.,
Refs.\onlinecite{RusKor-ent,KorJordan-undo,Ruskov-neg,prl-quadr,Beenakker,EngelLoss,Jordan-neg,Nori}).
Recent experiments on a single phase qubit provide an
interesting manipulation of its state
via negative-result measurement.\cite{Katz-Sci,Katz-Nature}
These advances support the hope that experimental implementation of
negative-result evolution is also possible in a DQD spin qubit system.

\section*{ACKNOWLEDGMENTS}

The authors would like to thank A.N. Korotkov, L.P. Pryadko,
and A.J. Rimberg
for fruitful discussions and remarks.
This work at Ames Laboratory was supported by the
Department of Energy --- Basic Energy
Sciences under contract No. DE-AC02-07CH11358.


\begin{thebibliography}{99}
\bibitem{Landauer}
     Independent tunneling events amount to the requirement of
     high detector voltage as mentioned in Sec. \ref{Hamiltonian_time_scales}.
     %
     Number of classical tunneling attempts per second can be evaluated as
     $A=\frac{2 e V}{2 \pi \hbar}$, consistent with the Landauer formula
     for average currents, given just after Eq.(\ref{int-2spin}).
\bibitem{EnsAverageView} From an ensemble averaged point of view, when the
              result of the QPC is ignored, any qutrit state will be ``frozen''
              for times $\Gamma^{-1} \lesssim t \ll \Gamma_{sw}$.
\bibitem{non-ideality-neg}
    Generally, we expect that possible presence of dephasing
    in Eq. (\ref{Bayes-upd2}), e.g. a factor $e^{-\gamma_{\mbox{\tiny eff}} t}$,
    will be strongly suppressed in the Zeno regime:
    $\gamma_{\mbox{\scriptsize eff}}\sim \Gamma_{sw}^2/\Gamma \ll \Gamma_{sw}$
    (compare with Ref.\cite{Kor-fidelity}).
\bibitem{improvement}
There is a room for
a factor of $2 - 10$ improvement of the charge sensitivity $\delta q$
(depending on the SET operational temperature\cite{Delsing})
until reaching the theoretical limit
obtained in Ref.\onlinecite{KorotkovPaalanen}
\end{thebibliography}
\end{document}